\documentclass[fleqn,usenatbib]{mnras}

\usepackage{newtxtext,newtxmath}
\usepackage{subfigure}

\usepackage[T1]{fontenc}
\usepackage{ae,aecompl}

\usepackage{graphicx}	
\usepackage{amsmath}	
\usepackage{amssymb}	
\usepackage{comment}
\usepackage[export]{adjustbox}
\usepackage{subfigure}

\title{Improving the Accuracy of Magnetic Field Tracing by Velocity Gradients: Principal Component Analysis}

\author[Hu, Yuen \& Lazarian]{
Yue Hu$^{1,2}$,
Ka Ho Yuen$^{2}$,\thanks{E-mail: kyuen2@wisc.edu}
A. Lazarian$^{2}$
\\
$^{1}$College of Electronics and Information Engineering, Tongji University, Shanghai, China\\
$^{2}$Department of Astronomy, University of Wisconsin-Madison, Madison, USA\\
}

\date{Accepted 3/7/2018. Received 18/4/2018; in original form.}

\pubyear{2018}

\begin{document}
\label{firstpage}
\pagerange{\pageref{firstpage}--\pageref{lastpage}}
\maketitle

\begin{abstract}
Tracing of the magnetic field with Velocity Gradient Technique (VGT) allows observers to probe magnetic field directions with spectroscopic data. In this paper, we employ the method of Principal Component Analysis (PCA) to extract the spectroscopic information most valuable for VGT. By using synthetic observation data from numerical simulations, we show that PCA acts in a way similar to spatial filtering along the velocity axis. We study both subsonic and supersonic simulations and show that with the PCA filtering the tracing of magnetic fields by the VGT is significantly improved. Using 21 cm GALFA data, we demonstrate that the PCA filtering improves the alignment of the velocity gradients and the Planck dust polarization.
\end{abstract}




\section{Introduction}
\label{sec:intro}

Turbulence is ubiquitous in the interstellar medium (ISM) at different scales \citep{AM95, CL10} and magnetic field plays an important role for most of the ISM physics. In particular, magnetic fields are essential for the star formation \citep{1976ApJ...210..326M,2015ApJ...808...48B,2013ApJ...770..151C}, propagation and acceleration of  cosmic ray \citep{1949PhRv...75.1169F,Sch10}, transport of heat and mass in the galaxy \citep{L06,NM01}. More recently, the importance of studying the structure of magnetic field was motivated by attempts to study elusive B-modes of cosmological origin \citep{2014JCAP...06..053F}. The latter produce polarization that is being confused with the foreground polarization arising from interstellar magnetic fields\citep{1989ApJ...346..728J,2016MNRAS.462.2343V}. However, the study of magnetic fields in the ISM is complicated. Therefore it is extremely interesting in finding alternative ways for magnetic field tracing.

The VGT technique employs either Velocity Centroid Gradients (VCGs) \citep{ GCL17,YL17a,YL17b} or Reduced Velocity Centroid Gradients (RVCGs) \citep{LY18a} or Velocity Channel Gradients (VChGs)\footnote{The technique is based on the theoretical \citep{LP00}. The theory predicts that the velocity caustics dominate the intensity fluctuations in thin channel maps.} (\citealt{LY18a}). In this paper, we use the VCGs, but the approach that we discuss in this paper is also applicable to other realizations of the VGT.

The VGT is founded by the modern understanding of  MHD turbulence theory (\citealt{GS95}, hereafter GS95) that includes the concept of fast turbulent reconnection (\citealt{LV99}, henceforth LV99) and is supported by numerical studies ( see \citealt{CV00,2001ApJ...554.1175M,CLV02, CL2003, K09}). Due to turbulent reconnection, motions perpendicular to the local direction of the magnetic field are not constrained, and therefore eddies rotating perpendicular to magnetic field have the Kolmogorov spectrum with eddy velocity $v_l\sim l_{\bot}^{1/3}$. The index $\bot$ in  $l_{\bot}$ indicate that the motions are perpendicular to the magnetic field of the eddy. As a result, the gradient of velocity scales as $v_l/l_{\bot}\sim l_{\bot}^{-1/3}$, means that the smallest resolved eddies induce the largest gradients. These gradients are perpendicular to the local direction of magnetic field. A more detailed explanation of the foundations of the VGT can be found in \citep{LY18a}. 

The VGT has been a fast developing branch of research. For instance, the tracing of the direction of magnetic field in diffuse \citep{YL17a} and self-gravitating media \citep{YL17b, LY18a} has been performed, as well as the estimations of the sonic ($M_s$, \citealt{YLL18}) and Alfvenic ($M_A$, \citealt{LYH18}) Mach numbers. As a separate development, the approach of studying magnetic fields with gradients has been also applied to synchrotron intensities\citep{Letal17}, which resulted in the Synchrotron Intensity Gradients (SIGs) technique, as well as to synchrotron polarization \citep{LY18b}, which resulted in two techniques, the Synchrotron Polarization Gradients (SPGs) and Synchrotron Polarization Derivative Gradients (SPDGs).\footnote{Incidentally, our approach of block averaging is also applicable to studies of gradients of column densities. The corresponding Intensity Gradient Technique (IGT) should not be confused with the Histograms of Relative Orientation (HRO) proposed in Soler et al. (2013). The IGT traces both magnetic fields and shocks (see \cite{YL17b}, \cite{LY18a}), while HRO provides a statistical relation between the relative orientation of the magnetic field and intensity gradients as a function of the column density. The latter is a measure calcuated for the entire image and it cannot be used to trace the spatial variations of magnetic fields. We view the IGT as a part of the gradient technique. Its synergy with the VGT was demonstrated e.g. in \cite{LY18a}}. The theoretical foundations of the procedures employed in the aforementioned techniques mentioned above are similar to those of the VGT, and therefore we expect that the improvements of the data analysis, in particular, the use of the Principal Component Analysis can be also advantageous for improving the accuracy of other gradient techniques, e.g. those dealing with synchrotron.

The practical application of VGT is affected by the quality of the data. The noise suppression method for VGT has been explored in \cite{Letal17} and elaborated in \cite{LY18a}, showing that a convolution of the observational map with a small $\sigma$ Gaussian kernel would retrieve the spatial structure of the molecular cloud.  Moreover, in \cite{YL17b} they showed that the filtering of non-turbulence contribution in Fourier space could improve the accuracy of VGT in tracing magnetic field. In this paper, we proceed with the work of improving magnetic field tracing with the VGT. For this purpose, we explore the application of the Principal Component Analysis (PCA). 

The PCA is widely used in image processing and image compression. Regarding astrophysical applications the PCA analysis was used in \cite{2002ApJ...566..276B,2002ApJ...566..289B} for obtaining the turbulence spectrum from observations. Later, in   \cite{2008ApJ...680..420H} the PCA was employed for studying turbulence anisotropies. Our present use of the PCA is different: we use it as a tool to provide the preliminary processing of the spectroscopic data. 

The idea of the PCA is that the image of size $N^2$ can be effectively represented by $n<N$ eigen-maps. The physical meaning of the eigenvalues from the PCA analysis is closely related to the value of the turbulence velocity dispersion $v^2$.   We apply the VCGs to the individual eigen-images and explore for which of them the magnetic field is traced the best. 

In what follows, we briefly describe the numerical code and setup for simulation in \S \ref{tab:sim}. In \S \ref{sec:res}, we test the implementation of the VCGs with the PCA using numerical simulations. \S \ref{sec:obs} shows the observational example with VCG-PCA technique. In \S \ref{sec:conclusion}, we give our discussion about our technique and conclusion.

\section{Numerical setting for synergistic use of PCA and VGT}
\label{tab:sim}
\begin{table}
\centering
\begin{tabular}{c c c c}
\hline
 Model Name & $M_S$ & $M_A$ & Resolution \\ \hline 
 Ms0.4Ma0.04 & 0.41 & 0.04 & $480^3$\\
 Ms0.8Ma0.08 & 0.92 & 0.09 & $480^3$\\
 Ms1.6Ma0.16 & 1.95 & 0.18 & $480^3$\\
 Ms3.2Ma0.32 & 3.88 & 0.35 & $480^3$\\
 Ms6.4Ma0.64 & 7.14 & 0.66 & $480^3$\\ 
\hline
\end{tabular}
\caption{\label{tab:sim} MHD simulations used in the present work.  $M_s$ and $M_A$ denote the instantaneous values of the  sonic and Alfven Mach numbers at each of the snapshots. }
\end{table}

For our studies of gradients with PCA, we use the same numerical cubes as in \cite{brazil18} (see Table \ref{tab:sim}). The simulation parameters with different combinations of Alfvenic Mach numbers $M_A=V_L/V_A$ and sonic Mach numbers $M_S=V_L/V_s$ are listed in Table \ref{tab:sim}. There $V_L$ is the turbulence injection velocity and $V_A$, and $V_s$ are the Alfven and sonic velocities, respectively. 

For this study, we consider the optically thin case ($\tau \sim \int \kappa(s) ds \ll 1$) and synthesize observational maps similar to that in \cite{brazil18} . We assume that the emissivity is proportional to density, but do not consider this as a significant limitation. For instance,  the case of emissivity proportional to the density squared is regarded in \citet{2017MNRAS.470.3103K} with the change of the results being insignificant. 

We denote the intensity within of the  Position-Position-Velocity (PPV) cubes as $\rho(x,y,v)$, and the cubes dimensions $n_x\times n_y\times n_v$, where the $n_v$ means the number of velocity channels along the spectral line direction v (line-of-sight direction, LOS), which is $n_v=400$ for our studies unless specifically mentioned. 

The PPV cubes are preprocessed first by the PCA similar to that described in \cite{2002ApJ...566..289B}. After that, the VCG technique is applied to the eigen-images of a different order. The product of PCA would be a set of eigen-images $I_i$ with decreasing order of eigenvalues $\lambda_i \sim v^2_i$, where the latter records the velocity variance along the line of sight. 

As we discussed in the \S \ref{sec:intro} that for studying turbulence the velocity variance is related to the eddy size along the line of sight. \cite{2008ApJ...680..420H} splits the PPV cube into vertical and horizontal Position-Velocity tires (PV tires), where every PV tire is a vertical or horizontal slice from the PPV map $\rho(x,y,v)$ averaged over the x-direction or y-direction, respectively. The eigenvalue is obtained by solving the eigenvalue equation for each PV tire.

\begin{figure*}
\centering
\subfigure[1st eigen-channel, AM=0.48
]
{\includegraphics[width=.3\linewidth,height=0.3\linewidth]{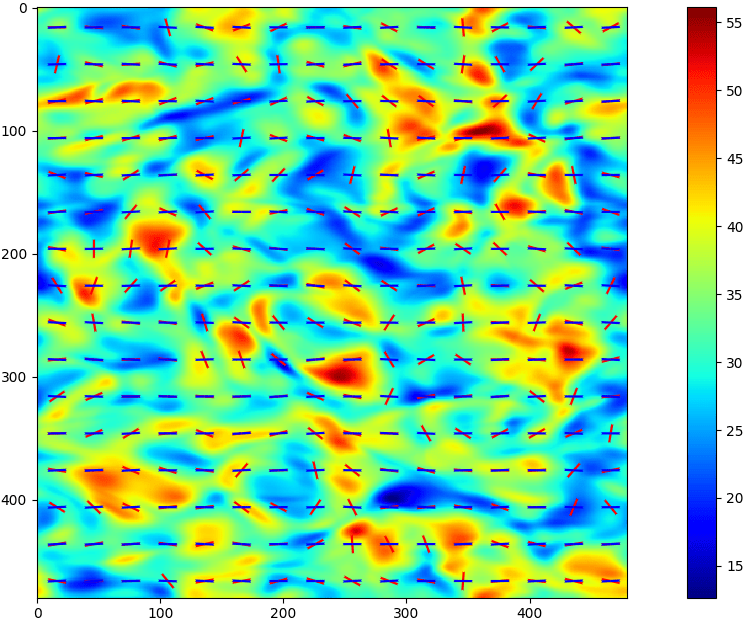}}
\centering
\subfigure[10th eigen-channel, AM=0.64
]
{\includegraphics[width=.3\linewidth,height=0.3\linewidth]{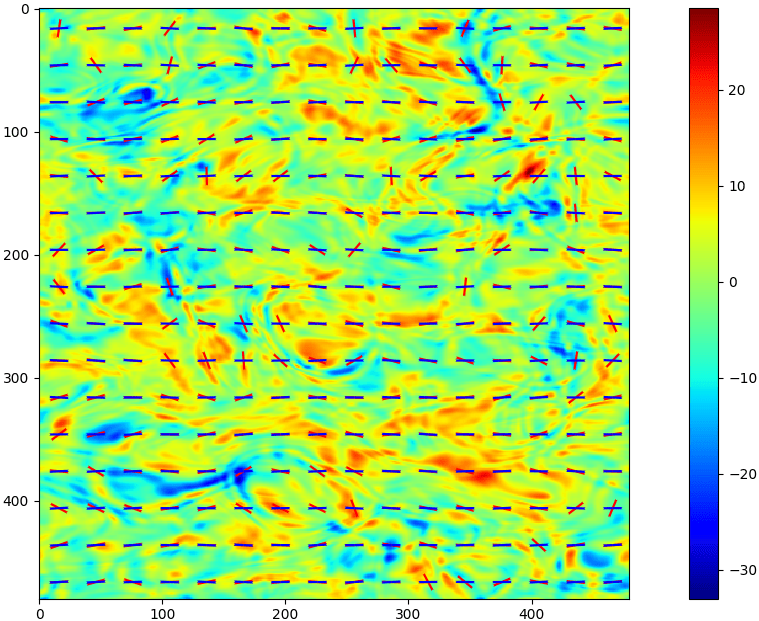}}
\centering
\subfigure[20th eigen-channel, AM=0.66
]
{\includegraphics[width=.3\linewidth,height=0.3\linewidth]{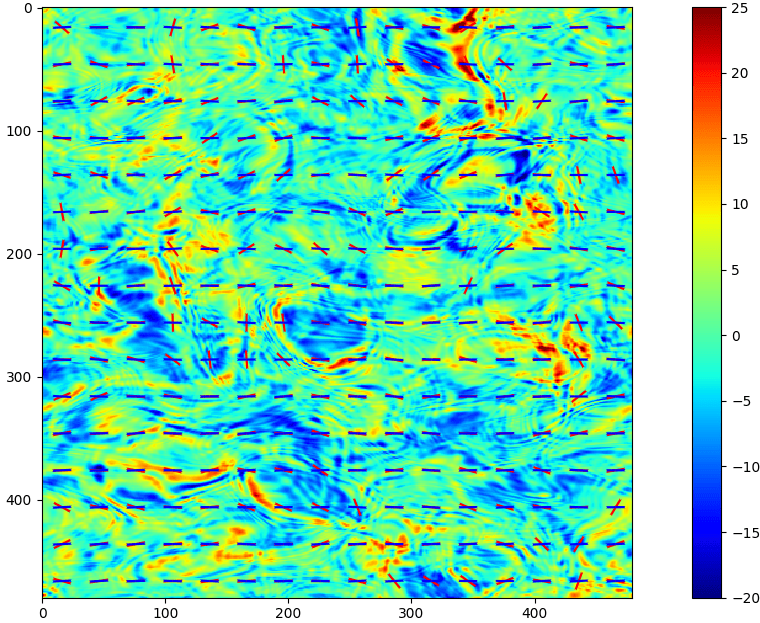}}
\centering
\subfigure[30th eigen-channel, AM=0.63
]
{\includegraphics[width=.3\linewidth,height=0.3\linewidth]{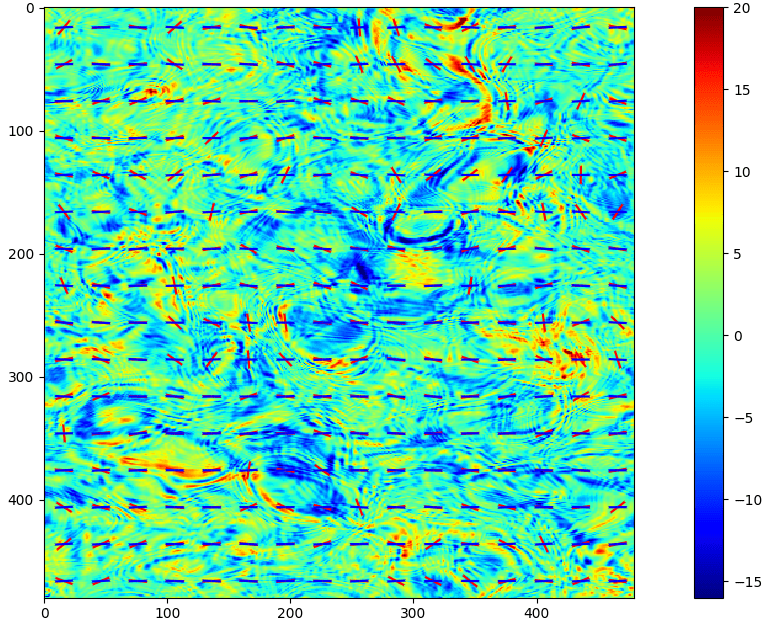}}
\centering
\subfigure[40th eigen-channel, AM=0.64
]
{\includegraphics[width=.3\linewidth,height=0.3\linewidth]{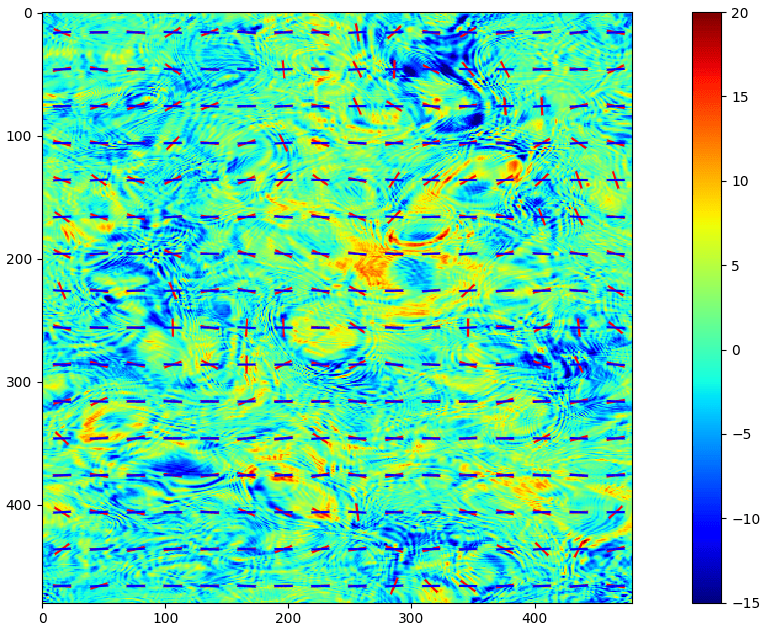}}
\centering
\subfigure[50th eigen-channel, AM=0.63
]
{\includegraphics[width=.3\linewidth,height=0.3\linewidth]{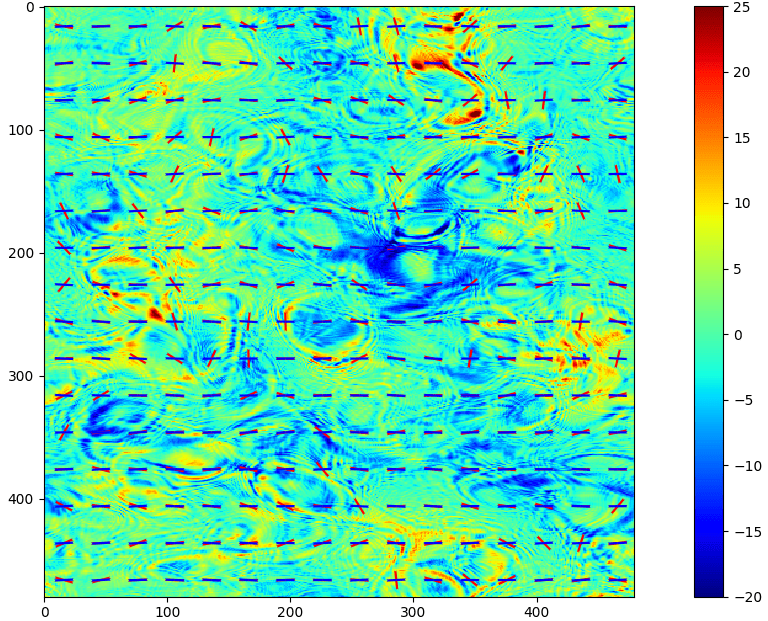}}
\caption{\label{fig:si-eigen0}The eigen-centroid maps with gradients (red) and magnetic field (blue) plotted with different eigenvalues. The simulation used here is Ms0.4Ma0.04 with $M_{S}=0.41$ and $M_{A}=0.04$. Please note, each figure is in individual color-scale.}
\end{figure*}

Similarly, in our work, we assume the PPV cube is properly normalized \footnote{In principle one shall use the normalized PPV cube $\rho' = \rho/\int \rho$. However, for the treatment of PCA, the difference of a constant does not alter the result. Therefore we stay with $\rho$ for simplicity.}, and treat the PPV cube $\rho(x,y,v)$ as the probability density function of three random variables $x,y,v$, we can then define the covariance matrix \citep{2002ApJ...566..276B} of each velocity channel  as: \footnote{The textbook definition of covariance matrix should be $S(v_1,v_2)=E(\rho(v_1) \rho(v_2))-E(\rho(v_1))E(\rho(v_2))$, where E is the expectation operator. However in  both \cite{2002ApJ...566..276B,2002ApJ...566..289B} and \cite{2008ApJ...680..420H} the second part is not included. In this work, we do not include this part either. However, we expect the inclusion of the second part brings only small effect to the eigenvalues of the covariance matrix if we are focusing only the largest eigenvalues.}
\begin{equation}
S(v_1,v_2) \propto \int dxdy \rho(x,y,v_1)\rho(x,y,v_2)
\end{equation}
hence an eigenvalue equation for this covariance matrix is: 
\begin{equation}
S\textbf{u}=\lambda\textbf{u}
\end{equation}
where the $\lambda_{i}$ are the eigenvalues associated with the eigenvectors $\textbf{u}_{i}$ with $i=1,2,...,n_v$. One can solve the eigenvalue equation to get the eigenvalue and eigenvector of each channel. The eigenvectors $\textbf{u}_{i}$ contain the weight of how one can construct the eigen-maps of rank $i$\footnote{ Here we are referring to the ordering index of eigenvalues from PCA after sorting them from the largest to smallest. } with the channel maps. We apply eigenvalues $\lambda_{i}$ as the weighting coefficients for each channel. Then the eigen-intensity maps $I_{eigen}$ and eigen-centroid maps $C_{eigen}$ can be computed by:
\begin{equation}
C_{eigen}(x,y)=\frac{\int dv\ \rho(x,y,v)\cdot v \cdot \lambda(v)}{I_{eigen}(x,y)}
\end{equation}
\begin{equation}
I_{eigen}(x,y)=\int dv\ \rho(x,y,v)\cdot \lambda(v)
\end{equation}

For the gradient computation, we shall follow the sub-block averaging method developed in \cite{YL17a}, which will tell the sub-block averaged orientation of gradients. The resultant gradients will be rotated $90^o$ to correspond to the expected magnetic field directions. The error estimation method \citep{LY18a} is also employed to signify how accurate the Gaussian fitting function used in sub-block averaging is when computing the average gradient direction within a sub-block. The orientation of gradients from VGT is compared with the synthetic polarization, assuming a constant emissivity in the dust grain alignment \citep{L07} . 

That means the mock Stokes parameters $Q(x,y)$ and $U(x,y)$\citep{2015PhRvL.115x1302C} can be expressed in terms of the angle $\theta$ between the x and y direction magnetic fields by $\tan\theta(x,y,z)=\frac{B_y(x,y,z)}{B_x(x,y,z)}$:
\begin{equation}
Q(x,y)\propto\int dz\rho(x,y,z)\cos(2\theta(x,y,z))
\end{equation}
\begin{equation}
U(x,y)\propto\int dz\rho(x,y,z)\sin(2\theta(x,y,z))
\end{equation}
The  polarization angle $\Phi=0.5 arctan2(\frac{U}{Q})$  is then defined correspondingly, which gives an probe of projected magnetic field in realistic scenarios.

The relative orientations between the $90^o$ rotated gradients and project magnetic field directions from polarization angles are measured by the \textbf{Alignment Measure (AM)} used in our previous studies \citep{GCL17, YL17a}: 
\begin{align}
AM=2(\langle cos^{2} \theta_{r}\rangle-\frac{1}{2})
\end{align}

Where $\theta_{r}$ is the relative angle between the gradients (rotated $90^o$) and the direction of the projected magnetic field. The range of AM is [-1,1]. When $AM = 1$, the gradients (rotated $90^o$) are parallel to the projected magnetic field. When $AM = -1$, the gradients (rotated $90^o$) are perpendicular to the projected magnetic field. We expect to get $AM\sim 1$ in most scenarios.
\begin{figure}
\centering
\includegraphics[width=0.99\linewidth]{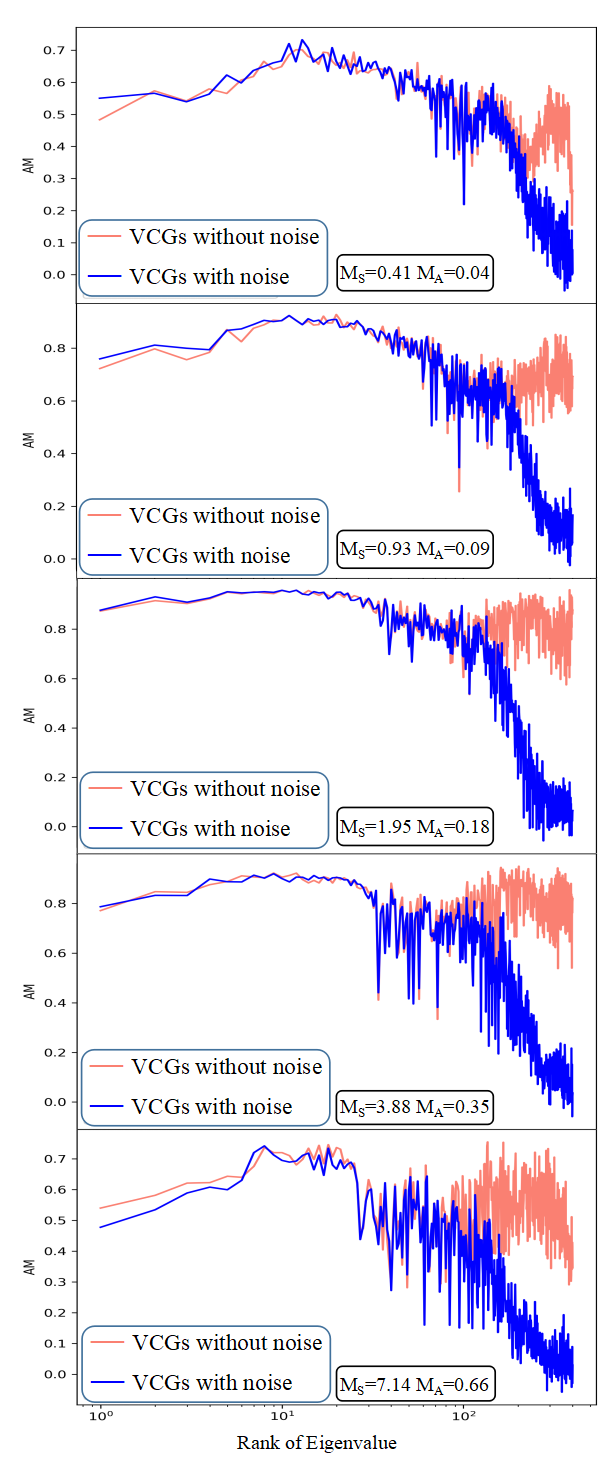}
\caption{\label{fig:pca-noiseC} Five plots showing the response of AM between gradients of eigen-centroids and projected magnetic field to the rank of eigenvalue (the maximum eigenvalue is ranked as the first one, the minimum eigenvalue is the last one) for both cases without noise (pink) and with noise added (blue).}
\end{figure}
\section{Applying VCGs to Eigen-images}
\label{sec:res}

The eigen-images produced by the PCA are the product to which we apply the VCGs analysis. For doing the latter, we apply the procedures described in our earlier papers, e.g. \citep{YL17a}, i.e. compute the sub-block averaged VCGs for each eigen-image and compare the obtained gradient directions with the projected magnetic field directions. Fig. \ref{fig:si-eigen0} illustrates the gradients and the structure of some selected eigen-centroids for the cube Ms0.4Ma0.04. One can see for the eigen-centroids, the first eigen-channel map shows a lower level of alignment, the rest are essentially equally aligned. The structure of the eigen-centroids becomes more filamentary when the rank of eigen-channel map increases.

We analyze the visual patterns in Fig \ref{fig:si-eigen0} using the AM-eigenvalue plot. The pink curves in Fig. \ref{fig:pca-noiseC} shows how the AM of the gradients from eigen-images and projected magnetic field varies concerning the eigenvalues from PCA analysis for the numerical cubes listed in Table \ref{tab:sim}. To test the power of PCA on noise reduction, we add white noise with mean amplitude $0.1 \sigma_C$ to the centroid maps. The results are shown as the blue curves in Fig. \ref{fig:pca-noiseC}. The $x$-axis in Fig. \ref{fig:pca-noiseC} represents the rank of eigenvalues sorted in decreasing order, i.e. if $\lambda_1>\lambda_2> ... >\lambda_n$, then we shall use the number 1 (the rank) to represent $\lambda_1$, rank 2 for $\lambda_2$ etc.  We see that for all simulations we tested, the peak rank is at around $\sim 10$. As the rank increases (i.e., smaller eigenvalues), the AM of the respective gradients of eigen-centroids to magnetic field decreases significantly. In noisy environments (blue curves in Fig. \ref{fig:pca-noiseC}), the AM of the images corresponding to the small ranks are approximately the same as the case without noise (pink curves), but the AM in higher rank cases drop significantly. The experiment in Fig. \ref{fig:pca-noiseC} shows that using the method of PCA before applying VGT, we can retrieve the strong signal part, which has a lower rank in PCA, from the noisy part, which has a higher rank.

\begin{figure}
\centering
\includegraphics[width=0.5\textwidth]{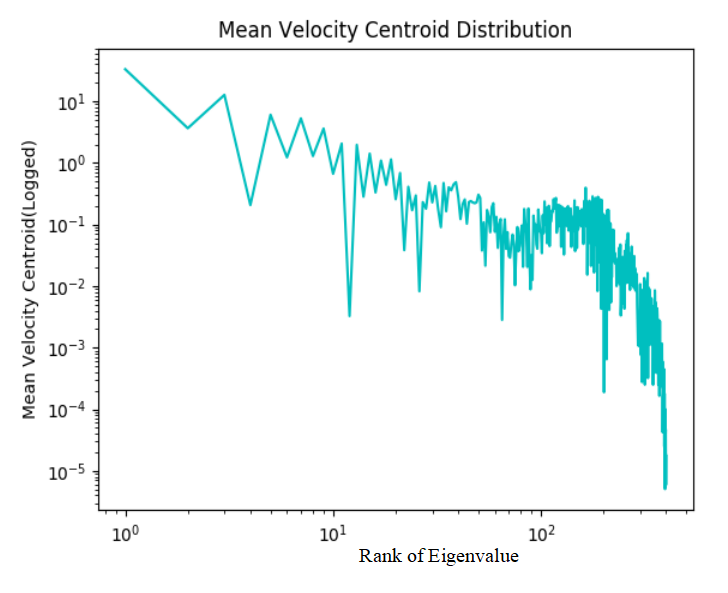}
\caption{\label{fig:meanIC} A plot showing how the eigen-centroid amplitudes varies with the rank of the eigen-values on the synthetic map from the cube Ms0.4Ma0.04.}
\end{figure} 

\begin{figure}
\centering
\includegraphics[width=0.48\textwidth]{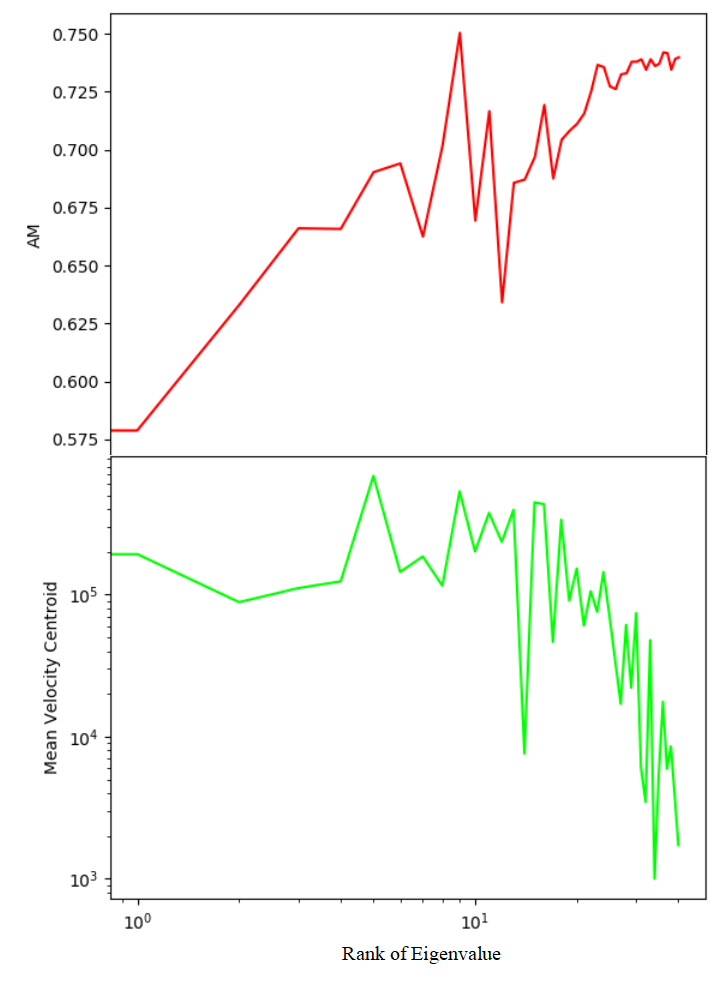}
\caption{\label{fig:susan_eigen} A plot showing the AM(top) and mean centorid amplitude(bottom) versus the rank of the eigen-values on the PPV cube from observation}
\end{figure}

\begin{figure*}
\centering
\includegraphics[width=0.98\textwidth]{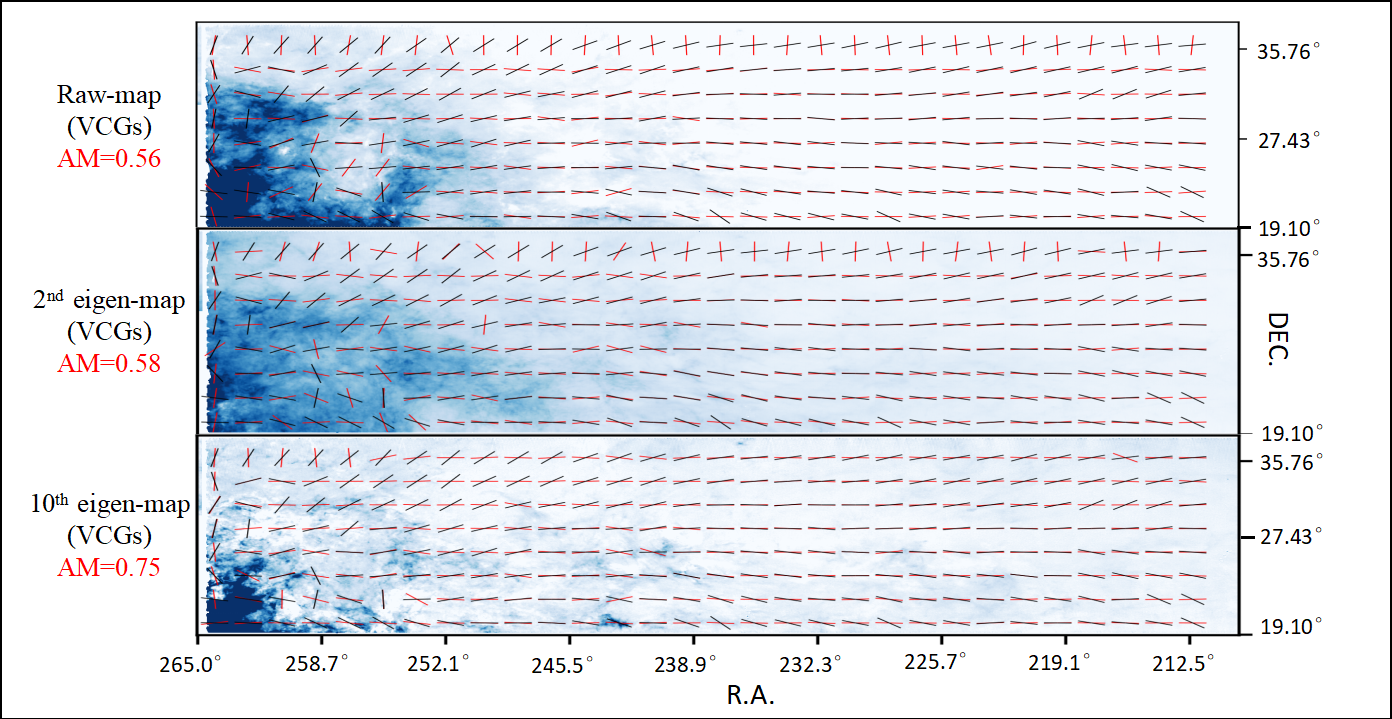}
\caption{\label{fig:susan} The region is from GALFA-HI and spans right ascension $212.5^{o}$ to $265^{o}$ and declination $19.1^{o}$ to $38.3^{o}$, stretches from b = $30^{o}$ above the Galactic plane to b =$81.7^{o}$, nearly Galactic zenith. We compare the gradients got from VCGs(red lines) with the Planck polarization data (black lines). Please note, each figure is in individual color-scale.}
\end{figure*}
We also notice that the AM for the images with ranks in the range $\sim 1-5$ is generally smaller than that for the ranks in the range $\sim 10-15$. The reason behind this is that the largest velocity dispersion $v^2$ extracted from PCA corresponds to the largest-scale eddies along the line of sight, that is affected by the energy injection.  Taking into account that for Alfvenic strong turbulence that $v^2\sim l^{2/3}$ (GS95, LV99), the images with range of ranks about $10-15$ correspond to the turbulent eddies in the inertial range of our numerical cubes ($k_{\rm inertial} \approx 10-30$). In fact, when we refer to the eigen-centroid amplitudes from the cube Ms0.4Ma0.04 (Fig \ref{fig:meanIC}), we can see the amplitude becomes insignificant after rank $>20$. The amplitude of the eigen-centroids with the rank higher than $20$ is at least $0.1 - 0.01$ compared to the first few eigen-centroids.  Therefore to use VCG$_{PCA}$ for its full potential, it is advantageous to remove the largest eigenvalues together with those having rank $>20$ to obtain the best result in magnetic field tracing.

\section{Application to observations}
\label{sec:obs}
For testing our recipe, we use the well-studied region from \cite{2015PhRvL.115x1302C}, with further information that can be found in \cite{susantail}. The region spans right ascension (R.A.) $212.5^{o}$ to $265^{o}$ and declination (DEC.) $19.1^{o}$ to $38.3^{o}$, covering a substantial piece of HI region with different physical conditions. The HI-cube has 41 velocity channels with each $\sim 3km/s$ wide. In previous studies \citep{YLL18,LYH18} we explored $M_s$ and $M_A$ in the same region, showing the region is super-sonic and sub-Alfvenic ($M_A\sim 0.75$), which is close to the condition we had in Table \ref{tab:sim}. We then use the same strategy as we did in \S \ref{sec:res} to analyze the gradient orientation with the PCA. 

We apply the PCA to the selected region and choose the $2^{nd}$ and $10^{th}$ eigen-centroid maps based on our experience that we had in Fig \ref{fig:pca-noiseC}. Since the PCA eigen-rank is similar to the wavenumber in the spectral analysis, the choice we made here should not be affected by the short numerical inertial range in our simulation as we are choosing the first few eigenvalues for analysis. We show the magnetic field tracing with VCGs and compare it to the magnetic field directions traced by the 353GHz Planck polarization data, which we illustrated in Fig.\ref{fig:susan} with two eigen-centroid maps $\lambda_2$ and $\lambda_{10}$. The corresponding figure showing the AM-eigenvalue variation is in Fig \ref{fig:susan_eigen}, which has the same trend as Fig \ref{fig:pca-noiseC}. The $10^{th}$ eigen-map has an obviously better AM compared to that of $2^{nd}$, which is consistent with the study we have in \S \ref{sec:res}.

\section{Discussion}

\subsection{Studying media magnetization}

The magnetization of the interstellar media can be characterized by the Alfven Mach number $M_A$. By itself, $M_A$ is critical parameter the knowledge of which is essential for understanding the vital astrophysical processes, including the transport and acceleration of cosmic rays (see \citealt{2014ApJ...784...38L}), transport of heat (see \citealt{L06}), etc. With known $M_A$ and known velocity dispersion one can get the value of the interstellar magnetic field (see \citealt{LYH18}). 

The technique of studying media $M_A$ using velocity gradients was suggested in \citet{LYH18}. This technique was tested with numerical simulations and applied to 21 cm data. The decrease of the noise that we observe applying the PCA technique for the initial filtering of the data is valuable for the studies of $M_A$. We plan to demonstrate this elsewhere. 

\subsection{Obtaining 3D structure of magnetic field}
The employment of eigenvalue decompositions through PCA also provides a way to study three-dimensional (3D) magnetic field. As we can isolate the contribution of turbulent eddies along the line of sight with PCA, we can then stack the prediction from different eigen-centroids and construct the 3D tomography by sorting the eigenvalue axis. In a separate development the gradients of synchrotron intensity \citep{Letal17} and polarization intensity \citep{LY18b} have been used to construct the 3D magnetic field morphology with multi-frequency measurements. A similar idea of constructing 3D magnetic field morphology with VGT on spectroscopic data has been tested in \cite{GL18} when the galactic rotation curve is available. With these 3D field tracing methods available, the productive application of VGT and synergy with different techniques will then shift the paradigm of studying magnetic fields from polarimetry measurements to studies of gradients on both interferometric and spectroscopic data.

\subsection{Application within other gradient techniques}

We expect the PCA filtering to be useful when applied with other velocity gradient techniques, e.g., with VChGs. However, the application of the procedure is not limited to the velocity gradients.

It is explained, e.g., in \cite{LY18b}, that the VGT is one of the techniques that employ the properties of MHD turbulence to study magnetic fields. Magnetic fluctuations enter Alfvenic turbulence in a symmetric way to velocity fluctuations. Therefore both synchrotron intensity gradients (see \citealt{Letal17}) and synchrotron polarization gradients \citep{LY18b} can be used to trace magnetic field and study $M_A$. Naturally, the improvement that we are suggesting here with the pre-filtering the images using the PCA seems an attractive possibility for these synchrotron-based techniques. 

We would like to mention that while the statistics of density fluctuations in MHD turbulence\citep{2005ApJ...624L..93B,2007ApJ...658..423K} does not follow closely, especially at high sonic Mach numbers, the statistics of velocity and magnetic field fluctuation, the Intensity Gradient Technique (IGT)\footnote{As we mentioned earlier, one should distinguish the IGT technique and the Histogram of Relative Orientation (HRO) \citep{2013ApJ...774..128S}. We would like to stress that the former employs the technology that we developed for the velocity and magnetic gradients and therefore can provide the spatial distribution of magnetic fields, shocks, regions of gravitational collapse, etc. (see \citealt{YL17b}, \citealt{LY18a}). It is important that the IGT technique does not require any polarization data to get this information. HRO, on the contrary, compares the relative orientation of the density gradients and the polarization directions as a function of column density.} are also very informative. We also expect to see the improvements for the IG technique when the PCA pre-filtering is employed. 

\section{Summary}
\label{sec:conclusion}

 In the present paper, we have utilized filtering of images using the PCA.  We have shown using both synthetic and observational maps, that the extraction of eigen-centroids with the rank number of $\sim 10$ can effectively probe the direction of magnetic field with a very high AM. As a result, for the studies of the projected magnetic field, the improved technique can provide higher accuracy of magnetic field tracing.

\section*{Acknowledgements}
AL acknowledges the support of NSF grants DMS 1622353, AST 1715754 and 1816234. This publication utilizes data from Galactic ALFA HI (GALFA HI) survey data set obtained with the Arecibo L-band Feed Array (ALFA) on the Arecibo 305m telescope. The Arecibo Observatory is operated by SRI International under a cooperative agreement with the National Science Foundation (AST-1100968), and in alliance with Ana G. Méndez-Universidad Metropolitana, and the Universities Space Research Association. The GALFA HI surveys have been funded by the NSF through grants to Columbia University, the University of Wisconsin, and the University of California. 








\label{lastpage}
\end{document}